%% file: cmt21_babbage.tex
\magnification=\magstep1 
\input boxedeps.tex
\input boxedeps.cfg
\SetepsfEPSFSpecial
\HideDisplacementBoxes
\font\bigbfont=cmbx10 scaled\magstep1
\font\bigifont=cmti10 scaled\magstep1
\font\bigrfont=cmr10 scaled\magstep1
\vsize = 23.5 truecm
\hsize = 15.5 truecm
\hoffset = .2truein
\baselineskip = 14 truept
\parskip = 3 truept
\topinsert
\centerline{IFUP-TH 60-97}
\vskip 3.0 truecm
\endinsert
\centerline{\bigbfont THE GROUND STATE OF MEDIUM-HEAVY NUCLEI}
\vskip 3 truept
\centerline{\bigbfont WITH NON CENTRAL FORCES}
\vskip 20 truept
\centerline{\bigifont Adelchi Fabrocini}
\vskip 8 truept
\centerline{\bigrfont Istituto Nazionale di Fisica Nucleare, Sezione di Pisa}
\vskip 2 truept
\centerline{\bigrfont and}
\vskip 2 truept
\centerline{\bigrfont Department of Physics, University of Pisa, Pisa, Italy}
\vskip 1.8 truecm

\centerline{\bf 1.  INTRODUCTION}
\vskip 12 truept

Correlated Basis Function (CBF) theory is a powerful 
tool to investigate the properties of many-body strongly interacting 
systems in several fields of physics. Its first succesful applications 
are found in homogeneous systems such as 
liquid Helium, electron systems (both in the forms of electron fluids and 
lattice structures) and infinite nuclear and neutron matter. The flexibility 
of CBF based approaches results in a realistic description 
of ground state (energy, momentum distribution, distribution functions 
and so on) as well as of dynamical (cross sections) quantities. 
While the g.s. is often better studied by MonteCarlo (MC)  
methods, CBF and perturbative theories in a CBF basis still provide 
one of the very few viable ways for an affordable, quantitative study of 
the response of many-body systems to external probes. 
In addition , the nuclear interaction and the induced nucleon-nucleon 
correlations  are strongly dependent on the states of the nucleons 
themselves, and the MC techniques adopted, for instance, in the case 
of liquid Helium g.s. are not easily extended to nuclear cases. 

As most of the world we know is not homogeneous, CBF addicted began in the 
last decade to extend the theory (and the connected technologies) to such 
interesting objects as Helium films and droplets and finite nuclear systems 
(usually named {\it nuclei}). In Helium the interaction is 
known and has a relatively simple form, so the 
difficulty lies in the large densities. On the contrary,  nuclei are 
considerably less dense and the actual problem is the aforementioned 
complicated structure of the interaction. Presently, there is a 
number of methods that exactly solve the light nuclei 
Schr\"odinger equation for realistic  hamiltonians. Faddeev [1,2],
Green Function Monte Carlo (GFMC) [3] and Correlated Hyperspherical 
Harmonics Expansion  [4] are used for 
the A=3,4 nuclei, and GFMC has been recently pushed up to A=7 [5]. 
Variational MonteCarlo (VMC) [6] methods has been also used in light nuclei 
and, if the spanned variational wave function space is large enough, then 
its description may be quite accurate 
(even if not exact). VMC has been extended to heavier nuclei, as 
$^{16}$O [7].

As the mass number, A, grows larger and the region of medium-heavy nuclei 
is approached, using MC with realistic hamiltonians becomes 
increasingly difficult and CBF theory  and cluster expansions represent 
an alternative and a competitive approach. In a series of papers, 
doubly closed shell nuclei (both in {\it ls} and {\it jj} coupling) were 
studied using the Fermi hypernetted chain (FHNC) [8] summation technique 
[9,10,11]. Semirealistic, central interactions and simple 
two-body correlations, depending on the interparticle distances and, 
at most, on the isospin of the correlated pair, were used and nuclei 
ranging from $^4$He to $^{208}$Pb were investigated. In a recent 
paper the same method has been extended to  interactions and correlations 
containing spin, isospin and tensor components for $^{16}$O and $^{40}$Ca 
nuclei, having doubly closed shells in {\it ls} coupling [12]. 
The review of the results obtained in this work will be the object of 
the present contribution. 

We consider a standard nuclear hamiltonian
$$
H=-{{\hbar^2}\over2\,m}\sum_i\nabla_i^2+\sum_{i<j}v_{ij}~~,\eqno(1)
$$
containing only a two-nucleon potential, $v_{ij}$, whose 
large interparticle distances behavior is dominated by meson exchange 
processes (one pion exchange, OPE). 
The intermediate and short distances parts of the potential are 
generally treated in semi-microscopic or purely phenomenological 
ways and several recipies are on the market. Their common feature is 
the accurate fit to the large body of available two nucleon scattering 
data up to energy $\sim 350$ MeV. 
For our purposes, we shall use a truncated version of the 
realistic Urbana $v_{14}$ model (U14) [13].
U14 is pametrized as the sum of 14 components, 
$$
v_{14,ij}=\sum_{p=1,14}v^p(r_{ij})O^p_{ij}~~,\eqno(2)
$$
with
$$
O^{p=1,6}_{ij}=
\left[ 1, {\bf \sigma}_i \cdot {\bf \sigma}_j, S_{ij} \right]\otimes
\left[ 1, {\bf \tau}_i \cdot {\bf \tau}_j \right]~~,\eqno(3)
$$
$S_{ij}$ is the  tensor operator and the remaining $p>6$ components contain linear 
and quadratic spin-orbit (${\bf L}\cdot{\bf S}$) and 
$L^2$ terms. The $p>6$ components are not retained in the 
$v_6$ (U6) truncation.

A CBF g.s. correlated A-nucleon wave function can be written 
as
$$
\Psi(1,2...A)=[ {\cal S}\prod_{i<j}F_{i,j} ]\Phi(1,2...A)~~,\eqno(4)
$$
{\it i.e.} a symmetrized product of two-body correlation operators, $F_{ij}$,  
acting on a mean field state,  $\Phi(1,2...A)$, given by a shell model 
wave function built up with  $\phi_\alpha(i)$ single particle wave 
functions. $F_{ij}$ is chosen of a form consistent with the interaction, 
$$
F_{ij}=\sum_{p=1,6}f^p(r_{ij})O^p_{ij}~~.\eqno(5)
$$
The correlation functions $f^p(r)$ are variational since they 
contain a set of parameters fixed by minimizing the g.s. 
expectation value of the hamiltonian, 
$\langle H \rangle=\langle \Psi|H|\Psi\rangle /\langle \Psi|\Psi\rangle$. 
$\langle H \rangle$ is expanded in Mayer-like cluster diagrams and 
infinite classes of these diagrams are summed by FHNC integral equations.
The cluster expansion method has been  widely applied to both finite  
and infinite, interacting systems with state independent 
(Jastrow) correlations . 

For a realistic approach to nuclear systems, the correlation operators 
are strongly state dependent and do not commute between each other. This fact 
prevents from the development of a complete FHNC theory for the correlated 
wave function of eq.(4). A single operator chain (SOC) approximation was 
adopted in ref.[14] for the operatorial ($p>1$) correlations, 
together with a a full FHNC treatment of the Jastrow ($p=1$) part 
in nuclear matter (NM). 
FHNC/SOC apparently  provides a satisfying description of infinite nuclear 
matter at saturation density and a neutron matter equation of state compatible 
with the neutron stars observational data [15].
However, no exact check for FHNC/SOC is presently available in infinite 
nucleon matter, apart the evaluation of some additional classes of diagrams. 
The estimated accuracy in the g.s. energy has been set to less than $1$ MeV  
at saturation [16,15]. 

We shall use FHNC/SOC theory to study the g.s. of $^{16}$O 
and $^{40}$Ca with the U6 interaction and the 
$^{16}$O results will be compared with cluster MC (CMC) results. 
In the CMC method the Jastrow contribution is exactly treated by MC 
sampling, and  that from the operatorial components is approximated 
by considering (via MC) up to four-body cluster terms. 
Higher order contributions are then extrapolated.

\vskip 28 truept
\centerline{\bf 2.  ONE- AND TWO-BODY DENSITIES AND}
\vskip 12 truept
\centerline{\bf THE EXPECTATION VALUE OF THE HAMILTONIAN}
\vskip 12 truept

The one- and two-body densities, 
$\rho_1({\bf r}_1)$ (OBD) and $\rho_2^p({\bf r}_1,{\bf r}_2)$ (TBD), enter 
(more or less explicitely) the evaluation of $\langle H \rangle$. 
They are defined as
$$
\rho_1({\bf r}_1)=
\langle \sum_{i} \delta({\bf r}_1 - {\bf r}_i) \rangle~~,\eqno(6) 
$$
and 
$$
\rho_2^p({\bf r}_1,{\bf r}_2)=
\langle \sum_{i\neq j} \delta({\bf r}_1 - {\bf r}_i) \delta({\bf r}_2 
- {\bf r}_j) 
O^p_{ij} \rangle ~~.\eqno(7)
$$
The FHNC theory for the TBD in presence of Jastrow correlations is 
described at length in ref.[9] and here it is given as granted. Let's just 
recall that it is written in terms of the Jastrow 
correlation, $f^J(r)=f^1(r)$, and of the {\it nodal} (or {\it chain}) and 
{\it elementary} (or {\it bridge}) functions, $N_{xy}({\bf r}_1,{\bf r}_2)$ 
and $E_{xy}({\bf r}_1,{\bf r}_2)$, representing the sums of the 
diagrams having those  topological structures, respectively. 
The $(xy)$ classification corresponds to the exchange character  
of the external points (1,2): $x(y)=d,e$ with $d=$direct (the point 
does not belong to any exchange loop), $e=$exchange  (the point does 
belong to a closed  exchange loop). 
Futhermore, $(xy=cc)$ ($c=$cyclic) diagrams are present, whose 
external points both belong to the same, non closed exchange loop.  

In presence of operatorial correlations, the nodal functions acquire 
a state dependence, $N^p_{xy}({\bf r}_1,{\bf r}_2)$. Because of the 
non commutativity of the correlations, the same topological diagrams  
may originate from several different orderings. Spin-isospin traces 
provide a weight to each of the diagrams, which, in turn, depends on 
the ordering itself. 
Keeping track of them is not feasible by FHNC-like integral 
equations and only selected classes of diagrams may be correctly 
summed. The SOC approximation consists in summing $p>1$ chains, 
where each link may contain just one operatorial element and Jastrow 
dressings at all orders. We remind that operatorial dependence comes 
also in account of the exchange of two nucleons, as the exchange operator is   
$P_{ij}^{ex}=-[1+O^2_{ij}+O^4_{ij}+O^5_{ij}]/4$.

The FHNC/SOC equations are given in ref.[12], where they are solved in 
the FHNC/0 approximation (corresponding to set to zero the 
elementary diagrams). The validity of this seemingly crude 
approximation will be discussed later.

 The structure of the OBD is 
$$
\rho_1({\bf r}_1)=\rho_1^J({\bf r}_1)\left[ 1 + U^{op}_d({\bf r}_1)\right] + 
U^{op}_e({\bf r}_1) C_d({\bf r}_1)~~,\eqno(8) 
$$
 with
$$
\rho_1^J({\bf r}_1)=\left[ \rho_0({\bf r}_1) + U^J_e({\bf r}_1)\right] 
C_d({\bf r}_1)~~.\eqno(9) 
$$
 $\rho_0({\bf r}_1)=\sum_\alpha \vert \phi_\alpha (1) \vert ^2$ is the 
mean field density, $C_d({\bf r}_1)=\exp \left\{ U^J_d({\bf r}_1)\right\}$,  
$U^J_{d(e)}({\bf r}_1)$ are Jastrow vertex corrections and  
$U^{op}_{d(e)}({\bf r}_1)$ represent operatorial vertex corrections, 
linked, by integral equations given in refs.[9,12],
to the FHNC nodal functions.

A measure of the accuracy of the computed OBD is the degree of fulfilment 
of its normalization, $S_1=A$, where 
$$
S_1=\int d^3r_1 \rho_1({\bf r}_1)~~.\eqno(10)
$$
In addition, the exact TBD has to comply with the pair saturation, $S_2=1$, and 
(for the nuclei we are considering) the isospin saturation, $S_{\tau}=-1$, 
properties, with 
$$
S_2={{1}\over{A(A-1)}}\int d^3r_1\int d^3r_2 
\rho_2^{p=1}({\bf r}_1,{\bf r}_2)~~,\eqno(11)
$$
and
$$
S_{\tau}={{1}\over{3A}}\int d^3r_1\int d^3r_2 
\rho_2^{p=4}({\bf r}_1,{\bf r}_2)~~.\eqno(12) 
$$
A spin saturation sum rule, $S_{\sigma}=-1$, holds only in absence of 
tensor correlations.

Deviations of the sum rules from their exact values are due to ({\it i}) the 
FHNC/0 scheme and ({\it ii}) the SOC approximation. In ref.[9] 
was found that $E_{ee}^{ex}$, {\it i.e.} the sum of the $ee$-elementary 
diagrams whose external points belong to the same exchange loop, 
may substantially contribute to both $S_\tau$ and to the potential energy, 
if the potential has large exchange terms. 

 In evaluating  $\langle H \rangle$, it is convenient to use the Jackson-Feenberg 
identity [17] for the kinetic energy, $\langle T \rangle$, with the result 
$$
\langle T \rangle = T_{JF} = T_\phi + T_F~~,\eqno(13)
$$
\topinsert
\vSlide{+2.5cm}
\centerline{\BoxedEPSF{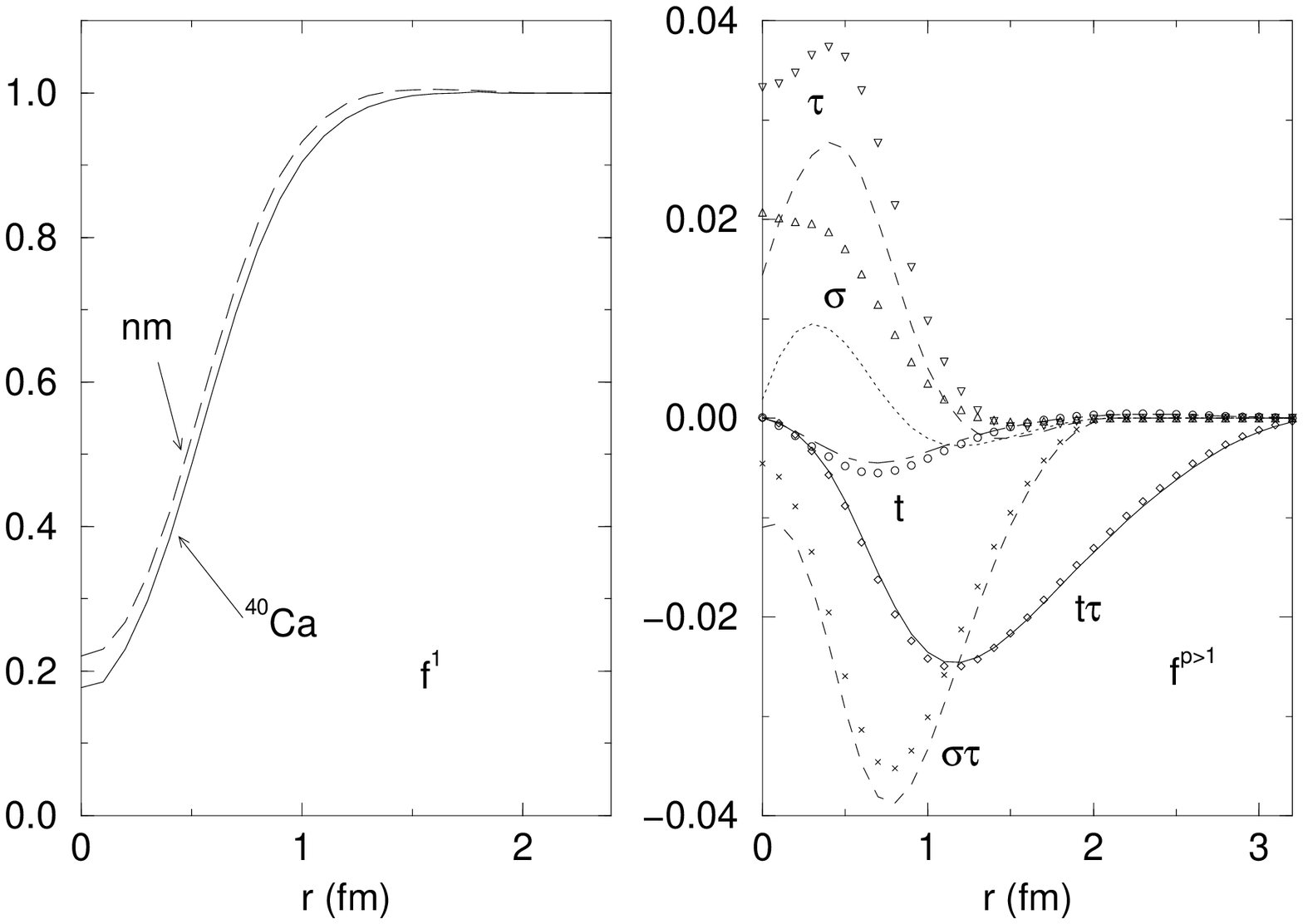 scaled 500}}
\endinsert
\noindent
{\bf Figure 1.}  Correlation functions for $^{40}$Ca and NM 
at saturation density. The operatorial correlations (right panel) are 
given as lines for $^{40}$Ca and symbols for NM.
\vskip 28 truept
\noindent
with 
$$
T_\phi=-A{{\hbar^2}\over{4m}}\langle \Phi^* G^2 \nabla_1^2 \Phi 
-  \left( \nabla_1 \Phi^*\right) G^2 \left( \nabla_1 \Phi \right) \rangle 
~~,\eqno(14)
$$
and 
$$
T_F=-A{{\hbar^2}\over{4m}}\langle \Phi^* 
\left[  G \nabla_1^2 G - \nabla_1 G \cdot  \nabla_1 G \right]
 \Phi \rangle~~,\eqno(15)
$$
where $G={\cal S}\prod F_{ij}$. Ref.[12] describes how to compute 
$T_\phi$ in FHNC/SOC. 

To evaluate $T_F$ and the two-body potential energy $\langle v \rangle=V_2$, 
we define an interaction energy $W=T_F+V_2=\langle H_{JF}\rangle$. The two-body 
operator, $H_{JF}^{ijk}(r_{12})$, is  
$$
H_{JF}^{ijk}(r_{12}) = - {{\hbar^2}\over{2m}}\delta_{j1}
\left\{ f^i(r_{12})\nabla^2 f^k(r_{12})-
 \nabla f^i(r_{12})\cdot\nabla f^k(r_{12})\right\}
$$
$$
 + f^i(r_{12})v^j(r_{12})f^k(r_{12})~~.\eqno(16)
$$

In FHNC/SOC, $W$ is split into four parts, $W=W_0+W_s+W_c+W_{cs}$,
where $W_0$ is the sum of the diagrams with only $p=1$, central chains 
between the interacting points (IP), connected by $H_{JF}$;  
$W_s$ sums diagrams having operatorial vertex corrections 
reaching the IP's and central chains; $W_c$ contains diagrams with one SOC 
between the IP's; $W_{cs}$ contains both. The lenghty formulas for 
$W$ are again found in ref.[12].

\vskip 28 truept

\centerline{\bf 3.  ENERGETICS}
\vskip 12 truept

 The results presented in this section have been obtained with 
the single particle wave functions, $\phi_\alpha(i)$, generated by 
a harmonic oscillator well with parameter $b$. 
\vfill\eject
\topinsert
\centerline{\bf{Table 1}}
\vskip 12 truept
\noindent
Breakup of the energy per nucleon, in MeV, for $^{16}$O with the U6 potential.
The J column gives the energies with the Jastrow corelated wave function, 
the O column those with the operatorial correlation and the CMC 
columns the cluster MonteCarlo results of ref.[18]. 
$E_{gs}=(\langle H \rangle - T_{cm})/A$.
\vskip 24 truept
\input  tables.tex
\nrows=11
\ncols=5
\begintable
~~~~~ & J & CMC(J) & O & CMC(O)  \crthick
$\langle v^1 \rangle$ | 0.88 | 0.93(28) | 2.33 | 2.35(43)~ \crnorule
$\langle v^2 \rangle$ | 1.25 | 1.27(08) | 1.97 | 2.00(13)~ \crnorule
$\langle v^3 \rangle$ | ~~~~ | ~~~~ | 0.25 | 0.27(01)~ \crnorule
$\langle v^4 \rangle$ | 2.40 | 2.43(12) | 2.29| 2.23(14)~ \crnorule
$\langle v^5 \rangle$ | -26.59 | -26.24(26) | -32.03 | -30.12(42)~ \crnorule
$\langle v^6 \rangle$ | ~~~~ | ~~~~ | -10.28 | -9.77(09)~ \cr
$\langle v \rangle$ | -22.07 | -21.56(25) | -35.47 | -33.03(31)~ \crnorule
$\langle T \rangle$ | 24.61 | 24.33(21) | 31.16 | 29.45(33)~ \cr
$\langle H \rangle$ |  2.54 | 2.77(09) | -4.33 | -4.59(10)~ \crnorule
$E_{gs}$ |  1.72 | ~~~~ | -5.15 | ~~~~~ \endtable
\vskip 14 truept
\endinsert
We kept $b$ fixed 
($b(^{16}O)=1.543~fm$ and $b(^{40}Ca)=1.654~fm$), even though it might 
be considered as a variational parameter. As far as the variational 
correlation, $F_{ij}$, is concerned, we follow ref.[9] and adopt 
an {\it Euler} correlation operator obtained by minimizing the 
energy at the second order of the cluster expansion, $\langle H_2 \rangle$. 
A corresponding choice was done in ref.[15] for NM. 
The variational parameters of the functions $f^p(r)$ are the {\it healing} 
distances, $d_p$, where $f^1(r\geq d_1)=1$, $f^{p>1}(r\geq d_p)=0$ 
and $(f^p)'(r=d_p)=0$. 
As in NM, we use only two healing distances: $d_c$ and 
$d_t$, for the central ($p=1,2,4,5$) and tensor ($p=3,6$) channels, 
respectively. An additional parameter is the quenching 
factor, $\alpha$, of the $p>1$ components of potential in the Euler equations. 
We have taken the U14 NM parameters, 
$d_c=2.15~fm$, $d_t=3.43~fm$ and $\alpha=0.8$, on the assumption that 
the short range behavior of the correlation does not vary drastically 
in going from the infinite to the finite systems. It is clear that a 
more accurate variational search for each nucleus would provide a 
lower ground state energy.

The $^{40}$Ca correlation functions are shown in Figure 1 and 
compared with the corresponding NM functions.  
They are similar, especially the longer ranged tensor parts. The 
most visible differences are found in the spin and isospin components 
and in the shortest range part of $f^1$. 

The error in the S$_1$ sum rule, in FHNC/0, is well less 
than $1\%$ in both nuclei for either the Jastrow (J) or the 
operatorial (O) correlations. The same accuracy is found for S$_2$ and
$S_{\tau}$ in the J-model after the insertion of the 
$E^{ex}_{ee}$ diagram (FHNC-1). The situation worsens for the 
O-model, where FHNC/SOC violates 
the sum rules by a maximum amount of $\sim 9\%$, as it 
was already found in NM [15].

The U6 energetics is displayed in Table 1 for $^{16}$O and 
Table 2 for $^{40}$Ca, for the J and O correlations. The Tables 
provide the potential, kinetic and total energy per nucleon. 
The separated expectation values of the components of the 
potential are also shown, together with the ground state energy, 
$E_{gs}$, obtained by subtracting the center of mass kinetic 
energy, $T_{cm}$, from $\langle H \rangle$. 

The $^{16}$O results are compared with the CMC calculations of Pieper 
[18]. FHNC shows an error of $1-2\%$ for $\langle T \rangle$ and 
$\langle v_2 \rangle$ in the J case. 
The total energy error is larger ($\sim 9\%$) as 
$\langle H \rangle$ is  given by the cancellation of two large numbers. 
The same situation is met in the O-model: the kinetic and potential energy 
errors are $5-7\%$. The absolute error in $\langle H \rangle$ is 
well less than 1 MeV, consistently  with the estimated accuracy of 
FHNC/SOC in NM [15]. The OPE parts of the potential provide 
most of the binding since, in its absence, $^{16}$O is not bound. 
The same holds in $^{40}$Ca, where 
the introduction of tensor correlations and potentials increases the 
kinetic energy by $\sim 5.6$ MeV, compensated by an additional 
potential energy contribution of $\sim -13.6$ MeV, providing a 
bound nucleus. 

In order to compare the theoretical energies with the experimental values 
one has to estimate the mean value of  the full U14 interaction. Its 
$p>6$ components represent the momentum dependent (MD) part of the 
potential, whose  expectation value is more difficult to evaluate than U6 
by means of cluster expansions. The usual strategy in NM is 
to include only low order (second and third) cluster MD contributions to
$\langle v^{MD} \rangle$. This procedure is probably accurate for U14, 
since its MD part contributes to the g.s. energy by less than 1 MeV at 
$\rho_{NM}=0.16$ fm$^{-3}$ ($\Delta E_{NM}^{MD}=0.44$ MeV), but it could be 
questionable for other potentials, as the recent Argonne $v_{18}$ model 
[19]. Moreover, it is known that a pure two-body 
interaction does not reproduce the NM saturation (energy, 
density and incompressibility) and the binding energies of light 
nuclei, so three-nucleon interactions, $v_{ijk}$, are to be introduced. The 
TNI model of ref.[13] approximated the effect of $v_{ijk}$ by adding 
two density dependent terms to U14: a repulsive TNR one, reducing the 
effect of the intermediate range part of $v_{ij}$, 
and an attractive TNA part. The TNR term was taken as 
$$
U14+TNR = \sum_{p=1,14}[v^p_\pi (r_{ij}) + 
v^p_I(r_{ij})~e^{-\gamma_1\rho_{NM}} + v^p_S (r_{ij})]
O^p_{ij}~~,\eqno(17)
$$
where $v^p_{\pi,I,S}$ are the long-, intermediate- and short-range 
parts of the potential, $\rho_{NM}$ is the NM density and 
$\gamma_1=0.15$ fm$^{-3}$. The TNA term was assumed to contribute as 
$$
TNA = 3 \gamma_2 \rho_{NM}^2 e^{-\gamma_3\rho_{NM}} ~~,\eqno(18)
$$
with $\gamma_2=-700$ Mev fm$^6$ and $\gamma_3=13.6$ fm$^3$. The values 
of the $\gamma$-parameters were obtained by fitting the NM saturation 
properties in FHNC/SOC.

Here we evaluate the TNR contribution by using 
$[\rho_1(r_i)\rho_1(r_j)]^{1/2}$ in place of $\rho_{NM}$ in (17) and 
$$
TNA = {1\over A} \int d^3r_1 \rho_1(r_1) \left[ 
3 \gamma_2 \rho_1^2(r_1) e^{-\gamma_3\rho_1(r_1)}\right] ~~.\eqno(19)
$$
\vfill\break
\topinsert
\centerline{\bf{Table 2}}
\vskip 12 truept
\noindent
As Table 1 for $^{40}$Ca. The CMC results are not available in this case.
\vskip 24 truept
\input  tables.tex
\nrows=11
\ncols=3
\begintable
~~~~~ & J & O  \crthick
$\langle v^1 \rangle$ | -1.41 | -0.21  \crnorule
$\langle v^2 \rangle$ | 1.57 | 2.30 \crnorule
$\langle v^3 \rangle$ | ~~~~ | 0.29 \crnorule
$\langle v^4 \rangle$ | 2.99 | 2.71  \crnorule
$\langle v^5 \rangle$ | -32.40 | -39.48  \crnorule
$\langle v^6 \rangle$ | ~~~~ | -14.14 \cr
$\langle v \rangle$ | -29.26 | -47.28 \crnorule
$\langle T \rangle$ | 30.55 | 39.69 \cr
$\langle H \rangle$ |  1.30 | -7.59 \crnorule
$E_{gs}$ |  1.01 | -7.87  \endtable
\vskip 14 truept
\endinsert

\vfill\break
\topinsert
\centerline{\bf{Table 3}}
\vskip 12 truept
\noindent
Energies per nucleon (in MeV) for $^{16}$O and $^{40}$Ca with 
the U14+TNI interaction. See text.
\vskip 24 truept
\input  tables.tex
\nrows=3
\ncols=8
\begintable
~~~~~ & $\langle H \rangle_{U6}$ & $\Delta E^{MD}$ & TNA 
      & $\langle H \rangle_{U14+TNI}$ & $\Delta E_c$ & $E_{gs}$ 
      & $E_{gs}^{expt}$ \crthick
$^{16}$O  | -1.88 | 0.33 | -4.53 | -6.08 | 0.05 | -6.85 | -7.72 \crnorule
$^{40}$Ca | -3.20 | 0.68 | -4.59 | -7.11 | 0.05 | -7.35 | -8.30  \endtable
\vskip 14 truept
\endinsert
$\Delta E^{MD}$ is estimated in local density approximation (LDA) as
$$
\Delta E^{MD} = {1\over A} \int d^3r_1 \rho_1(r_1) 
\Delta E^{MD}_{NM}[ \rho_1(r_1)] ~~.\eqno(20)
$$

Table 3 gives the $^{16}$O and $^{40}$Ca energies with the U14+TNI 
interaction and the operatorial correlation model. $\Delta E_c$ is the 
Coulomb energy correction taken from ref.[11]. The experimental 
ground state binding energies  per nucleon  are underestimated by 
$\sim 1$ MeV in both nuclei. The MD contribution is enough 
low to makes us enduring the LDA in its evaluation. We are considerably 
satisfied with these results, most af all in consideration of the 
further energy lowering that could be obtained by a careful variational 
minimization and of the FHNC/SOC accuracy.

\vskip 28 truept

\centerline{\bf 4.  CONCLUSIONS}
\vskip 12 truept

In this contribution we have presented some results for the ground state 
energy of doubly closed shell nuclei in {\it ls} coupling ($^{16}$O and 
$^{40}$Ca) in the framework of the correlated basis function theory. 
The correlated wave function includes central and tensor correlations, whose 
non commutativity does not allow for a complete FHNC treatement,  
contrary to the purely scalar case. We have extended the  
single operator chain approximation scheme (as used in realistic 
studies of nuclear and neutron  matter) to the finite nuclei. 
The U6 truncation of the realistic Urbana $v_{14}$ nucleon-nucleon potential 
has been adopted as a test.

By the analysis of the sum rules we have shown that FHNC/SOC 
provides an accurate one-body density. A comparably  good accuracy 
is obtained for the normalizations of the two-body density ($S_2$ 
and $S_\tau$) when tensor correlations are not included. Their  
insertion slightly worsens the excellent fullfiment of $S_{2,\tau}$, 
but the maximum violation is $\sim 9~\%$, similar to what was 
found in nuclear matter. 

The comparison of the energy contributions in $^{16}$O for the U6 
interaction with the cluster Monte Carlo results shows a 
maximum disagreement varying from $\sim 2~\%$ 
for the Jastrow model to $\sim 7~\%$ for the tensor model.  
The absolute error in the ground state energy per 
nucleon is well less 1 MeV, compatible with 
the estimated accuracy of the FHNC/SOC approach in nuclear matter at 
saturation density.  

The ground state expectation value of the complete realistic U14+TNI 
interaction has been evaluated by estimating the small momentum 
dependent contribution in local density approximation. The binding 
energies are about 1 Mev lower than the experimental values. 
Further improvements are reasonably to be expected by a better 
variational minimization. To our knowledge and opininion, this is 
one of the very few reliable microscopic evaluations of the g.s. energy of 
these nuclei with a realistic hamiltonian. Recently derived nuclear 
potentials provide a higher quality fit to the nucleon-nucleon 
scattering data, however their momentum dependence is stronger 
than in U14. Consequently, a better than LDA treatment of this 
part is imperative if they are to be used in the nuclei we have 
studied. Our group is working along this line, as well as to the 
insertion of explicit three-nucleon forces.

\vskip 28 truept

\centerline{\bf ACKNOWLEDGMENTS}
\vskip 12 truept

It is a pleasure to aknowledge the constant and fruitful collaboration 
with F.Arias de Saavedra, G.Co', S.Fantoni and P.Folgarait. We thank 
the Institute for Nuclear Theory at the University of Washington for 
its hospitality and the Department of Energy for its partial support 
during the completion of this work.

\vskip 28 truept

\centerline{\bf REFERENCES}

\vskip 12 truept

\item{[1]}
C. R. Chen, G. L. Payne, J. L. Friar and B. F. Gibson, 
Phys.\ Rev.\  {\bf C3}, 1740 (1986). 
\smallskip

\item{[2]}
A. Stadler, W. Gl\"ockle and P. U. Sauer, Phys.\ Rev.\  {\bf C44}, 2319 (1991).
\smallskip

\item{[3]}
J. Carlson, Phys.\ Rev.\  {\bf C38}, 1879 (1988).
\smallskip

\item{[4]}
A. Kievsky, M. Viviani and S. Rosati, 
Nucl.\ Phys.\  {\bf A551}, 241 (1993).
\smallskip

\item{[5]}
B. S. Pudliner, V. R. Pandharipande, J.  Carlson, S. C. Pieper 
 and R. B. Wiringa, Phys.\ Rev.\  {\bf C56}, 1720 (1997).
\smallskip

\item{[6]}
R. B. Wiringa, Phys.\ Rev.\  {\bf C43}, 1585 (1991).
\smallskip

\item{[7]}
S. C. Pieper, R. B. Wiringa and V. R. Pandharipande, 
Phys.\ Rev.\  {\bf C46}, 1741 (1992).
\smallskip

\item{[8]}
S. Rosati, in {\it From nuclei to particles}, 
Proc. Int. School E. Fermi, course LXXIX, ed. A. Molinari (North Holland,
Amsterdam, 1982).
\smallskip

\item{[9]}
G. Co', A. Fabrocini, S. Fantoni and I.E. Lagaris, 
Nucl.\ Phys.\ {\bf A549}, 439 (1992).
\smallskip

\item{[10]}
G. Co', A. Fabrocini and S. Fantoni, 
Nucl.\ Phys.\  {\bf A568}, 73 (1994).
\smallskip

\item{[11]}
F. Arias de Saavedra, G. Co', A. Fabrocini and S. Fantoni, 
Nucl.\ Phys.\  {\bf A605}, 359 (1996).
\smallskip

\item{[12]}
A. Fabrocini, F. Arias de Saavedra, G. Co' and P. Folgarait, 
Pisa Preprint IFUP-TH 40/97.
\smallskip

\item{[13]}
I. E. Lagaris and V. R. Pandharipande, 
Nucl.\ Phys.\  {\bf A359}, 331 (1981); {\bf A359}, 349 (1981).
\smallskip
 
\item{[14]}
V. R. Pandharipande and R. B. Wiringa, Rev.\ Mod.\ Phys.\ 
{\bf 51}, 821 (1979).
\smallskip

\item{[15]}
R. B. Wiringa, V. Ficks ad A. Fabrocini, Phys.\ Rev.\ {\bf C38}, 1010 (1988).
\smallskip

\item{[16]}
R. B. Wiringa, Nucl.\ Phys.\  {\bf A338}, 57 (1980).
\smallskip

\item{[17]}
S. Fantoni and S. Rosati, Phys.\ Lett.\  {\bf B84}, 23 (1979).
\smallskip

\item{[18]}
S. C. Pieper, private communication.
\smallskip

\item{[19]}
R. B. Wiringa, V. G. J. Stoks and R. Schiavilla, 
Phys.\ Rev.\  {\bf C48}, 646 (1995).
\smallskip

\end

%% file: tables.tex
%
\newbox\hdbox%
\newcount\hdrows%
\newcount\multispancount%
\newcount\ncase%
\newcount\ncols
\newcount\nrows%
\newcount\nspan%
\newcount\ntemp%
\newdimen\hdsize%
\newdimen\newhdsize%
\newdimen\parasize%
\newdimen\spreadwidth%
\newdimen\thicksize%
\newdimen\thinsize%
\newdimen\tablewidth%
\newif\ifcentertables%
\newif\ifendsize%
\newif\iffirstrow%
\newif\iftableinfo%
\newtoks\dbt%
\newtoks\hdtks%
\newtoks\savetks%
\newtoks\tableLETtokens%
\newtoks\tabletokens%
\newtoks\widthspec%
%
%
\immediate\write15{%
CP SMSG GJMSINK TEXTABLE --> TABLE MACROS V. 851121 JOB = \jobname%
}%
%
%
\tableinfotrue%
\catcode`\@=11
%
%
\def\tstrut{\vrule height3.1ex depth1.2ex width0pt}%
\def\and{\char`\&}
\def\tablerule{\noalign{\hrule height\thinsize depth0pt}}%
\thicksize=1.5pt
\thinsize=0.6pt
\def\thickrule{\noalign{\hrule height\thicksize depth0pt}}%
\def\ctr#1{\hfil\ #1\hfil}%
%
%
%
%
\tablewidth=-\maxdimen%
\spreadwidth=-\maxdimen%
\def\tabskipglue{0pt plus 1fil minus 1fil}%
%
%
\centertablestrue%
%
%
%
%
\parasize=4in%
\gdef\ARGS{########}
\gdef\headerARGS{####}
\def\@mpersand{&}
{\catcode`\|=13
\gdef\letbarzero{\let|0}
\gdef\letbartab{\def|{&&}}%
\gdef\letvbbar{\let\vb|}%
}
{\catcode`\&=4
\def\ampskip{&\omit\hfil&}
\catcode`\&=13
\let&0
\xdef\letampskip{\def&{\ampskip}}%
\gdef\letnovbamp{\let\novb&\let\tab&}
}
\def\begintable{
   \begingroup%
   \catcode`\|=13\letbartab\letvbbar%
   \catcode`\&=13\letampskip\letnovbamp%
   \def\multispan##1{
      \omit \mscount##1%
      \multiply\mscount\tw@\advance\mscount\m@ne%
      \loop\ifnum\mscount>\@ne \sp@n\repeat%
   }
   \def\|{%
      &\omit\widevline&%
   }%
   \ruledtable
}
\long\def\ruledtable#1\endtable{%
%
%
%
   \offinterlineskip
   \tabskip 0pt
   \def\widevline{\vrule width\thicksize}
   \def\endrow{\@mpersand\omit\hfil\crnorm\@mpersand}%
   \def\crthick{\@mpersand\crnorm\thickrule\@mpersand}%
   \def\crnorule{\@mpersand\crnorm\@mpersand}%
   \let\nr=\crnorule
   \def\endtable{\@mpersand\crnorm\thickrule}%
   \let\crnorm=\cr
%
%
   \edef\cr{\@mpersand\crnorm\tablerule\@mpersand}%
   \the\tableLETtokens
%
%
   \tabletokens={&#1}
%
%
   \countROWS\tabletokens\into\nrows%
   \countCOLS\tabletokens\into\ncols%
%
%
   \advance\ncols by -1%
   \divide\ncols by 2%
   \advance\nrows by 1%
%
%
   \iftableinfo %
      \immediate\write16{[Nrows=\the\nrows, Ncols=\the\ncols]}%
   \fi%
%
%
   \ifcentertables
      \ifhmode \par\fi
      \line{
      \hss
   \else %
      \hbox{%
   \fi
      \vbox{%
         \makePREAMBLE{\the\ncols}
         \edef\next{\preamble}
         \let\preamble=\next
         \makeTABLE{\preamble}{\tabletokens}
      }
      \ifcentertables \hss}\else }\fi
   \endgroup
   \tablewidth=-\maxdimen
   \spreadwidth=-\maxdimen
}
\def\makeTABLE#1#2{
   {
   \let\ifmath0
   \let\header0
   \let\multispan0
%
%
   \ncase=0%
   \ifdim\tablewidth>-\maxdimen \ncase=1\fi%
   \ifdim\spreadwidth>-\maxdimen \ncase=2\fi%
   \relax
%
   \ifcase\ncase %
      \widthspec={}%
   \or %
      \widthspec=\expandafter{\expandafter t\expandafter o%
                 \the\tablewidth}%
   \else %
      \widthspec=\expandafter{\expandafter s\expandafter p\expandafter r%
                 \expandafter e\expandafter a\expandafter d%
                 \the\spreadwidth}%
   \fi %
   \xdef\next{
      \halign\the\widthspec{%
      #1
      \noalign{\hrule height\thicksize depth0pt}
      \the#2\endtable
%
      }
   }
   }
   \next
}
\def\makePREAMBLE#1{
   \ncols=#1
   \begingroup
   \let\ARGS=0
   \edef\xtp{\widevline\ARGS\tabskip\tabskipglue%
   &\ctr{\ARGS}\tstrut}
   \advance\ncols by -1
   \loop
      \ifnum\ncols>0 %
      \advance\ncols by -1%
      \edef\xtp{\xtp&\vrule width\thinsize\ARGS&\ctr{\ARGS}}%
   \repeat
   \xdef\preamble{\xtp&\widevline\ARGS\tabskip0pt%
   \crnorm}
   \endgroup
}
\def\countROWS#1\into#2{
   \let\countREGISTER=#2%
   \countREGISTER=0%
   \expandafter\ROWcount\the#1\endcount%
}%
\def\ROWcount{%
   \afterassignment\subROWcount\let\next= %
}%
\def\subROWcount{%
   \ifx\next\endcount %
      \let\next=\relax%
   \else%
      \ncase=0%
      \ifx\next\cr %
         \global\advance\countREGISTER by 1%
         \ncase=0%
      \fi%
      \ifx\next\endrow %
         \global\advance\countREGISTER by 1%
         \ncase=0%
      \fi%
      \ifx\next\crthick %
         \global\advance\countREGISTER by 1%
         \ncase=0%
      \fi%
      \ifx\next\crnorule %
         \global\advance\countREGISTER by 1%
         \ncase=0%
      \fi%
      \ifx\next\header %
         \ncase=1%
      \fi%
      \relax%
      \ifcase\ncase %
         \let\next\ROWcount%
      \or %
         \let\next\argROWskip%
      \else %
      \fi%
   \fi%
   \next%
}
\def\counthdROWS#1\into#2{%
\dvr{10}%
   \let\countREGISTER=#2%
   \countREGISTER=0%
\dvr{11}%
\dvr{13}%
   \expandafter\hdROWcount\the#1\endcount%
\dvr{12}%
}%
\def\hdROWcount{%
   \afterassignment\subhdROWcount\let\next= %
}%
\def\subhdROWcount{%
   \ifx\next\endcount %
      \let\next=\relax%
   \else%
      \ncase=0%
      \ifx\next\cr %
         \global\advance\countREGISTER by 1%
         \ncase=0%
      \fi%
      \ifx\next\endrow %
         \global\advance\countREGISTER by 1%
         \ncase=0%
      \fi%
      \ifx\next\crthick %
         \global\advance\countREGISTER by 1%
         \ncase=0%
      \fi%
      \ifx\next\crnorule %
         \global\advance\countREGISTER by 1%
         \ncase=0%
      \fi%
      \ifx\next\header %
         \ncase=1%
      \fi%
\relax%
      \ifcase\ncase %
         \let\next\hdROWcount%
      \or%
         \let\next\arghdROWskip%
      \else %
      \fi%
   \fi%
   \next%
}%
{\catcode`\|=13\letbartab
\gdef\countCOLS#1\into#2{%
   \let\countREGISTER=#2%
   \global\countREGISTER=0%
   \global\multispancount=0%
   \global\firstrowtrue
   \expandafter\COLcount\the#1\endcount%
   \global\advance\countREGISTER by 3%
   \global\advance\countREGISTER by -\multispancount
}%
\gdef\COLcount{%
   \afterassignment\subCOLcount\let\next= %
}%
{\catcode`\&=13%
\gdef\subCOLcount{%
   \ifx\next\endcount %
      \let\next=\relax%
   \else%
      \ncase=0%
      \iffirstrow
         \ifx\next& %
            \global\advance\countREGISTER by 2%
            \ncase=0%
         \fi%
         \ifx\next\span %
            \global\advance\countREGISTER by 1%
            \ncase=0%
         \fi%
         \ifx\next| %
            \global\advance\countREGISTER by 2%
            \ncase=0%
         \fi
         \ifx\next\|
            \global\advance\countREGISTER by 2%
            \ncase=0%
         \fi
         \ifx\next\multispan
            \ncase=1%
            \global\advance\multispancount by 1%
         \fi
         \ifx\next\header
            \ncase=2%
         \fi
         \ifx\next\cr       \global\firstrowfalse \fi
         \ifx\next\endrow   \global\firstrowfalse \fi
         \ifx\next\crthick  \global\firstrowfalse \fi
         \ifx\next\crnorule \global\firstrowfalse \fi
      \fi
\relax
      \ifcase\ncase %
         \let\next\COLcount%
      \or %
         \let\next\spancount%
      \or %
         \let\next\argCOLskip%
      \else %
      \fi %
   \fi%
   \next%
}%
\gdef\argROWskip#1{%
   \let\next\ROWcount \next%
}
\gdef\arghdROWskip#1{%
   \let\next\ROWcount \next%
}
\gdef\argCOLskip#1{%
   \let\next\COLcount \next%
}
}
}
\def\spancount#1{
   \nspan=#1\multiply\nspan by 2\advance\nspan by -1%
   \global\advance \countREGISTER by \nspan
   \let\next\COLcount \next}%
\def\dvr#1{\relax}%
\def\header#1{%
\dvr{1}{\let\cr=\@mpersand%
\hdtks={#1}%
\counthdROWS\hdtks\into\hdrows%
\advance\hdrows by 1%
\ifnum\hdrows=0 \hdrows=1 \fi%
\dvr{5}\makehdPREAMBLE{\the\hdrows}%
\dvr{6}\getHDdimen{#1}%
{\parindent=0pt\hsize=\hdsize{\let\ifmath0%
\xdef\next{\valign{\headerpreamble #1\crnorm}}}\dvr{7}\next\dvr{8}%
}%
}\dvr{2}}
\def\makehdPREAMBLE#1{
\dvr{3}%
\hdrows=#1
{
\let\headerARGS=0%
\let\cr=\crnorm%
\edef\xtp{\vfil\hfil\hbox{\headerARGS}\hfil\vfil}%
\advance\hdrows by -1
\loop
\ifnum\hdrows>0%
\advance\hdrows by -1%
\edef\xtp{\xtp&\vfil\hfil\hbox{\headerARGS}\hfil\vfil}%
\repeat%
\xdef\headerpreamble{\xtp\crcr}%
}
\dvr{4}}
\def\getHDdimen#1{%
\hdsize=0pt%
\getsize#1\cr\end\cr%
}
\def\getsize#1\cr{%
\endsizefalse\savetks={#1}%
\expandafter\lookend\the\savetks\cr%
\relax \ifendsize \let\next\relax \else%
\setbox\hdbox=\hbox{#1}\newhdsize=1.0\wd\hdbox%
\ifdim\newhdsize>\hdsize \hdsize=\newhdsize \fi%
\let\next\getsize \fi%
\next%
}%
\def\lookend{\afterassignment\sublookend\let\looknext= }%
\def\sublookend{\relax%
\ifx\looknext\cr %
\let\looknext\relax \else %
   \relax
   \ifx\looknext\end \global\endsizetrue \fi%
   \let\looknext=\lookend%
    \fi \looknext%
}%
%
%
\def\tablelet#1{%
   \tableLETtokens=\expandafter{\the\tableLETtokens #1}%
}%
\catcode`\@=12